# Transition in High and Low Temperature Superconductors as Bounded Bose Einstein Condensation

V.Y. Butko[a,b]

[a]Ioffe Physical Technical Institute, Russian Academy of Science (RAS), 26 Polytechnichesaya, St. Petersburg 194021, Russia.

[b]St. Petersburg Academic University, Nanotechnology Research and Education Center, RAS, 8/3 Khlopin, St. Petersburg, 195220, Russia.

**The superconducting transition in High Temperature Superconductors(HTS) has not been understood[1]. Traditional Bose Einstein Condensation(BEC) theory provides $T_c$ estimates that are many times higher than the experimental values[2,3]. The proposed model of Bounded Bose Einstein Condensation(BBEC) assumes that the energy of a superconducting carrier in condensate is bounded by Fermi energy. BBEC theory predicts $T_c$ from the energetic balance that neglects any carrier pairing term and assumes 3 dimensional (3D) superconducting transport in Low Temperature Superconductors(LTS), and 2D superconducting transport channeled near the nodes of the Fermi arcs in HTS. BBEC produces $T_c$ scaling relations that are approximated by semi-empirical laws[4,5], and provides $T_c$ in agreement with experimental data. This agreement in HTS shows the role of BBEC as the main mechanism of superconducting transitions, indicates transport channeling (possibly related to the stripes[6-8]), and demonstrates a minor contribution into the energetic balance from pairing or pairing above $T_c$ (possibly related to the pseudogap[9-11]).**

For an ideal 3D gas of Bose particles the temperature of BEC is given by[3] $T_c=(3.31\hbar^2 n_s^{2/3})/(m^* k_B)$ (here here $n_s$ is the density of coherent superconducting carriers, ħ is the reduced Planck constant, $m^*$ is mass of the particle). This expression calculated with typical in HTS parameters predicts[3] a value of $T_c$~3000K that is many times higher than the experimental values. Therefore the interpretation of the Uemura law as a phenomenon related to BEC[2] while being intuitively appealing has not been quantitatively confirmed. Another BEC related[2] insightful approach for explanation of Uemura law is based on the estimate of the phase stiffness to fluctuations in superconductors[12]. This approach limits the scope of the theory mainly to HTS. Particularly the obtained upper bound on $T_c$($T_\theta^{max}$) in $UBe_{13}$ is equal[12] to 270 K, while $T_c$ in this material is[2] 0.9K, in Pb: $T_\theta^{max}$ is[12] 1.4 $10^6$ K, while[5] $T_c$~7.2K. Uemura and Holmes laws[4,5], however, are observed not only in HTS but also in LTS, indicating a more general nature of these phenomena.

I propose that the disagreement between BEC predictions and experiments is due to the difference in energy between a single boson particle which has zero kinetic energy when its momentum is zero and a Cooper pair which may have kinetic energy even when its momentum is zero due to an internal degree of freedom.This model focuses on general aspects of superconductivity as a quantum coherent phenomena and leaves the specific charge carrier pairing mechanism out of the scope. The following main assumptions are made.

1. Superconducting carriers are pairs of fermions and $n_{ns}$=$n_{nn}$-$n_s$, where $n_{nn}$ is the density of normal carriers in the normal phase, $n_{ns}$ is the density of incoherent (normal) carriers in the superconducting phase.
2. The transition from a superconducting phase to a normal phase due to a temperature increase occurs at $T_c$ at which ($U_n$-$U_s$) is balanced by the term in Helmholtz Free energy, ($T_c(S_n$-$S_s)$), and $T_c$ is estimated by the expression $T_c=(U_n-U_s)/(n_s(S_n-S_s))$, where ($U_n$-$U_s$) and ($S_n$-$S_s$), correspondingly, are differences in internal energy and in entropy between the normal and the superconducting phases.

3. $S_n$-$S_s$ can be calculated as $k_B\ln(\Omega_n/\Omega_s)$, where $\Omega_s$ is the number of microstates that can be occupied by a superconducting  carrier in the superconducting phase, $\Omega_n$ is the number of microstates that can be occupied by this carrier after it became normal above $T_c$ , and $k_B$ is Boltzmann constant.
4. $U_n$- $U_s$ can be obtained by a summation of the corresponding differences in energy ($\delta E$) of the individual charge carriers, which change their state through the superconducting transition. This theory takes into account that normal charge carriers as fermions occupy individual quantum states, and therefore occupy an extra volume in the momentum space compared to coherent superconducting carriers that as bosons can occupy the same quantum state. This extra volume and the corresponding to it difference in internal energy between the normal phase and the superconducting phase ($U_n$-$U_s$) is determined by the deviation of momentum of a normal carrier from its average value due to quantum uncertainty (Quantum Deviation (QD)).
5. The Fermi energy in the normal phase and in the superconducting phase can be measured or calculated, using QD similar to Sommerfeld theory of normal metals[13], as $E_F(n_{nn})=p_{Fnorm}^2/2m$ and $E_F(n_{ns})=p_{Fsuper}^2/2m$, correspondingly (here $p_{Fnorm}=p_F(n_{nn})$ and $p_{Fsuper}=p_F(n_{ns})$ are Fermi momentums in the normal phase and in the superconducting phase, correspondingly, m is the mass of a free electron.
6. QD is of the same order of magnitude as quantum uncertainty in momentum that is given by Heisenberg uncertainty relation $\Delta p_{xQU} \geq \hbar/(2\Delta x)$, and QD is estimated by $\Delta p$ $=\Delta p_x \sim w\hbar/\Delta x$, where $\Delta p_x$ is QD of a carrier momentum in x direction, $\Delta x$ is uncertainty in the carrier position, $w \approx 1/2$ and $\Delta x \approx 6L$ are used, where L is a mean free path of a charge carrier.

7. $\delta E$ can be estimated as $E_F(n_{nn})-E_F(n_{ns})$. In fact: the highest energy occupied by a normal carrier in the normal phase is equal to $E_F(n_{nn})$. To estimate $\delta E$ one needs additionally to estimate the energy of the superconducting carriers. Superconducting pairs are bosons that are coherent to each other and can occupy the same quantum state. Therefore they occupy a negligible volume in the momentum space compared to the volume occupied by normal carriers. However, superconducting carriers are not coherent to normal carriers and the internal kinetic energy of a superconducting pair with zero total momentum is bounded from below by a sum of energies of the paired carriers. The above $\delta E$ estimate is obtained for superconducting pairing between carriers that have opposite momentums and energies close to $E_F(n_{ns})$ similar to Bardeen Cooper Schrieffer(BCS) theory.

To explore significance of BBEC ($U_n$-$U_s$) term in the energetic $T_c$ balance we take into account this term and neglect any additional terms related to details of Cooper pairing in a particular superconductor. Making the summation (assumption #4) in its integral form, the following estimate can be obtained. (See also Fig.1a, Fig.1b).

$T_c = (1/n_s)\int_0^{n_s}\big(E_F(n_{nn}) - E_F(n_{nn}-n_s)\big)dn_s\,/\,(k_B \ln(\Omega_n\,/\,\Omega_s))$, here

$\delta E = E_F(n_{nn}) - E_F(n_{nn}-n_s) = (2\delta p\; p_F(n_{nn}-n_s) + \delta p^2)/(2m)$,

where $\delta p = p_F(n_{nn}) - p_F(n_{nn}-n_s)$.

Assumption #6 allows to estimate $\delta p$ in many important cases from the following equations. $V_{ps}=n_s\,(w\hbar)^j/2$, here $V_{ps}$ is the volume in the momentum space that according to assumptions #5 and #6 corresponds to the normal carrier density equal to $n_s$, j is the dimensionality of the space, the factor ½ is due to 2 possible spin orientations.

For spherically symmetrical 3 dimensional Fermi surface (Fig.1a)

$p_F(n_{nn3D}) \approx ((3/8\pi)\, n_{nn3D})^{1/3} w\hbar$,

$\delta p \approx (3/8\,\pi)^{1/3} n_{s3D} w\hbar/((n_{nn3D})^{2/3} + n_{nn3D}^{1/3}(n_{nn3D}-n_{s3D})^{1/3} + (n_{nn3D}-n_{s3D})^{2/3})$,

where $n_{s3D}$ is 3 dimensional superfluid density, obtained from $n_{s3D}/m = c^2/(4\pi\,\lambda^2 e^2)$, $\lambda$ is London penetration depth.

$S_n$-$S_s$ can be estimated by $k_B \ln(\Omega_n/\Omega_s) \approx k_B \ln(\delta p/\Delta p)$,(Fig.1a),where $\delta p/\Delta p > 1$.

Using simplifying condition, $n_{s3D} << n_{nn3D}$, obtain $\delta E \approx (1/2m) 2 p_F(n_{nn3D}) \delta p$ and

$T_c \approx 0.01 n_{s3D}\, (\hbar)^2/(n_{nn3D}{}^{1/3} m k_B \ln(L n_{s3D}/(n_{nn3D})^{2/3}))$ . (1)

For hole doped cuprate HTS as established by Angle Resolved Photoemission Spectroscopy (ARPES)[14-18] Fermi surface forms arcs one of which is schematically shown in Fig.1b. BBEC takes into account only charge carriers located inside the closed regions of the 2D momentum plane (see the hatched region in Fig.1b) that are bordered by the Fermi arcs and by the circle centered around zero momentum point with $p_{Fnorm}$ radius. These regions according to BBEC approach contain normal charge carriers in the normal phase and are depopulated at temperatures below $T_c$ due to the condensation bounded by Fermi arcs (this picture is supported by ARPES observations[14,16] of the broad and long Fermi arc regions in the normal phase and the sharp and short Fermi arc regions in the superconducting phase in HTS). The exact shape of these regions varies in HTS, however, under simplified assumptions of the arc radius equal to $p_{Fnorm}$ in these regions and of a small $\delta p(\delta p << p_F)$ the estimate, $\delta p_{max} \approx n_{s2D}{}^{2/3}(w\hbar)^{4/3}/(4 p_{Fnorm}{}^{1/3})$, can be obtained (see Fig.1b), where $n_{s2D}$ is 2 dimensional superfluid density, $n_{s2D} = n_{s3D}/N$, here N is the number of copper oxygen planes per unit length in the direction that is perpendicular to these planes. For underdoped, optimally doped and moderately overdoped HTS $S_n$-$S_s \approx k_B \ln(r_{arc}\delta p_{max}/(2\Delta p))$(here factor ½ is the estimate of a ratio of $\delta p_{max}$ to the average $\delta p$ in the regions, $r_{arc} \approx 1.4$(see figures[14,16])is the ratio of the arc length in the normal phase to the arc length in the superconducting phase,$(r_{arc}\delta p_{max}/(2\Delta p))>1$). Using $\delta E \approx (1/2m)2p_{Fnorm}\delta p_{max}$, obtain:

$$T_c \approx (0.06/m)\, n_{s2D}^{2/3} p_{Fnorm}^{2/3} \hbar^{4/3} / (k_B \ln(0.84 L n_{s2D}^{2/3} \hbar^{1/3} / p_{Fnorm}^{1/3})) \qquad (2)$$

Formula(2) is calculated, using $p_{Fnorm} \approx 1.2 p_{Fsuper}$ and $p_F = k_F\, \pi\hbar/a$, here $k_F$, is dimensionless Fermi momentum, $a \approx 0.38$nm is the inplane lattice constant.

In Fig. 2 we have plotted values of $T_c$ calculated by formula (2) and the values of $T_c$ measured in the HTS (see Table1) versus values given by formula (2). In this plot the best agreement between BBEC theory and the experiments would corresponds to the smallest deviation of the experimental data from the calculated straight line. The left inset in Fig. 2 illustrates similar comparison between predictions of the formula (1) and the $T_c$ values measured in Ta, Pb, and Nb (see Table1). The right inset in Fig.2 illustrates deviation of the experimental values of $T_c$ from BBEC estimates. Using the experimental data[19], we have produced in Fig. 3 a normalized phase diagram versus Sr doping for $La_{2-x}Sr_xCuO_4$ HTS by plotting the experimentally measured $T_c$, $T_c$ calculated by formula (2), and $T_c$ given by Uemura law. As one can see the general trend of all 3 curves are similar, but BBEC predictions have smaller deviation $(\sigma = (\sum_i (T_{cexpi} - T_{ctheori})^2/s)^{1/2}$, i is the measurement number and s is the number of measurments) from the experimental data compared to Uemura law. The future theory may include specific pairing mechanisms and more accurate estimates of quasiparticle velocities[20]. However, even in present form

BBEC estimates better agree with experimental results compared to BEC theory[2,3].

BBEC theory provides a universal approach to calculation of dependence of $T_c$ on the superfluid density and on other parameters. The obtained Eq.(1) and Eq.(2) predict scaling relations in isotropic 3D materials, $T_c \sim n_{s3D}/(n_{nn3D}^{1/3}\ln(n_{s3D}L/(n_{nn3D})^{2/3}))$,

and in hole doped HTS,

$T_c \sim n_{s2D}^{2/3}p_{Fnorm}^{2/3}/\ln(An_{s2D}^{2/3}L/p_{Fnorm}^{1/3})$,(here $A=0.84\hbar^{1/3}$),

that are approximated by semi-empirical Uemura($T_c \sim n_s$) and Homes($T_c \sim n_s/\sigma$, here $\sigma$ is the conductivity)laws[4,5].

The observed good agreement between BBEC predictions and the experimental results is evidence that BBEC is the main mechanism of the superconducting transition in a number of HTS and LTS.

**Acknowledgment**

This work was supported in part by the Russian Foundation for the Basic Research, by the St. Petersburg Scientific Center of the RAS, and by Presidium of the RAS. The author thanks A.V. Butko for his help.

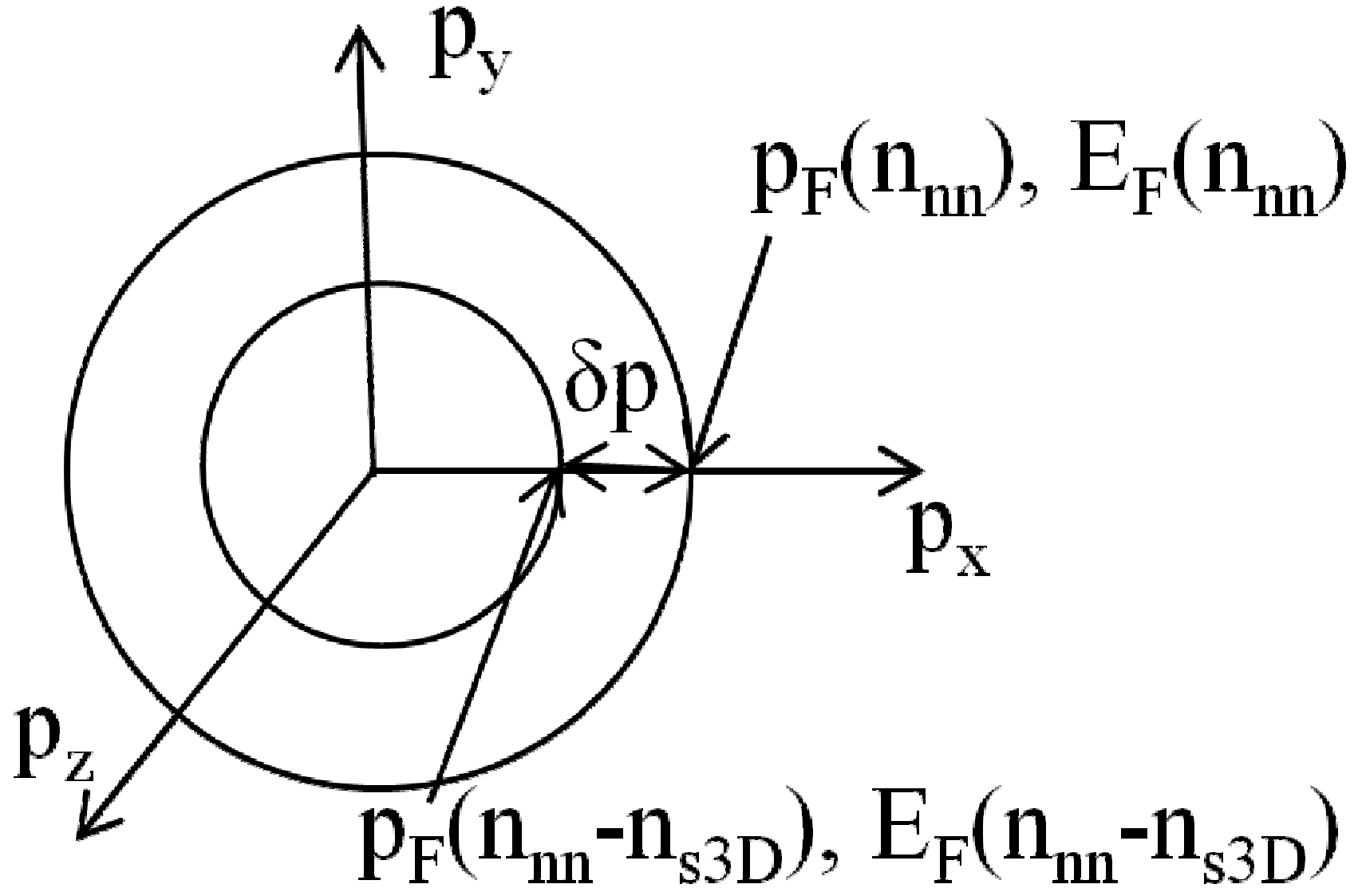

Figure 1a

A schematic view of 3 dimensional spherical Fermi surface.

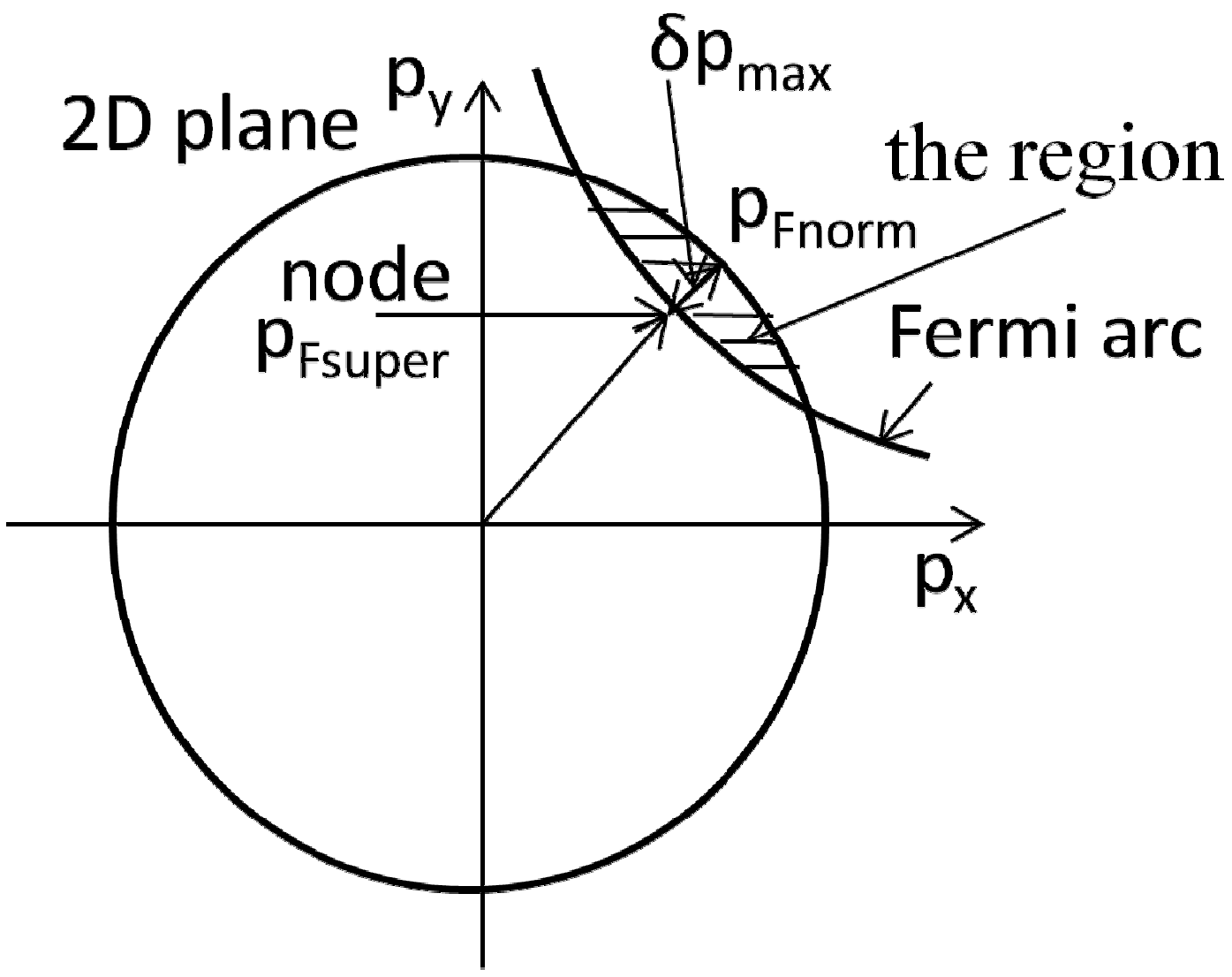

Figure 1b

A schematic view of 2 dimensional arc Fermi surface in HTS.

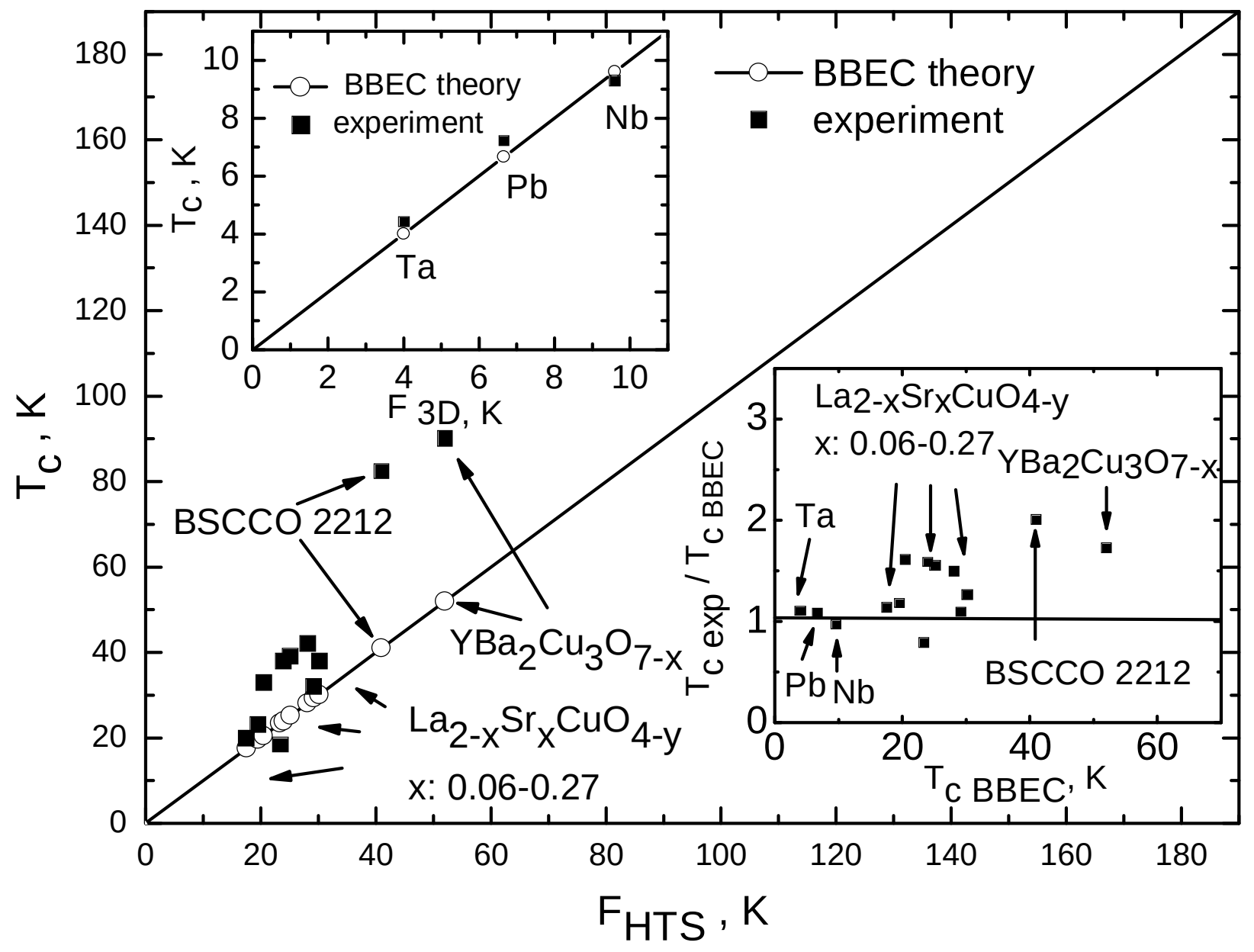

Figure 2. $T_c$ measured in HTS and $T_c$ given by Eq. (2) versus $F_{HTS}$ given by Eq. (2). The left inset: $T_c$ measured in LTS and $T_c$ given by Eq. (1) versus $F_{3D}$ given by Eq. (1). The right inset: the experimental $T_c$ normalized to BBEC $T_c$ versus BBEC $T_c$.

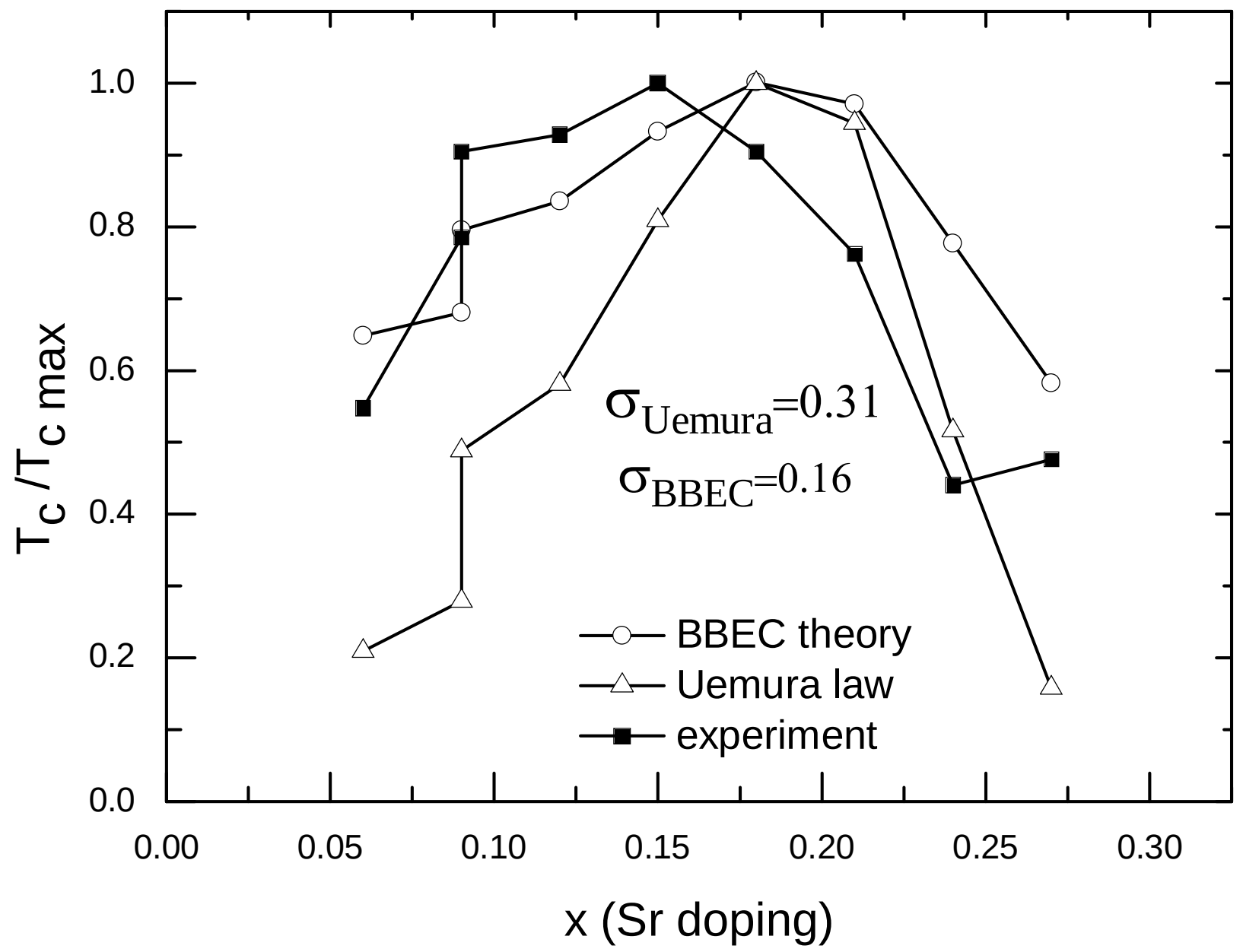

Figure 3. Phase diagrams of the normalized $T_c$ versus Sr doping in $La_{2-x}Sr_x CuO_4$.

Table 1. The used parameters

| Material | $\lambda$, nm | $k_F$, | $n_{nn3D}$, $cm^{-3}$ | $n_{s2D}$, $cm^{-2}$ | L, nm | $T_{c\ exp}$,K |
|---|---|---|---|---|---|---|
| $La_{1.94}Sr_{0.06}CuO_4$ | 471[19] | 0.435[14] | | 8,42 $10^{12}$ | 40.5 [&] | 23[19] |
| $La_{1.91}Sr_{0.09}CuO_4$ | 408[19] | 0.430[14] | | 1,12 $10^{13}$ | 40.5 [&] | 33[19] |
| $La_{1.91}Sr_{0.09}CuO_4$ | 309[19] | 0.430[14] | | 1,96 $10^{13}$ | 40.5 [&] | 38[19] |
| $La_{1.88}Sr_{0.12}CuO_4$ | 283[19] | 0.426[14] | | 2,33 $10^{13}$ | 40.5 [&] | 39[19] |
| $La_{1.85}Sr_{0.15}CuO_4$ | 240[19] | 0.422[14] | | 3.24 $10^{13}$ | 40.5 [#] | 42[19] |
| $La_{1.82}Sr_{0.18}CuO_4$ | 216[19] | 0.417[14] | | 4 $10^{13}$ | 40.5 [&] | 38[19] |
| $La_{1.79}Sr_{0.21}CuO_4$ | 222[19] | 0.413[14] | | 3.79 $10^{13}$ | 40.5 [&] | 32[19] |
| $La_{1.76}Sr_{0.24}CuO_4$ | 300[19] | 0.409[14] | | 2.08 $10^{13}$ | 40.5 [&] | 18.5[19] |
| $La_{1.73}Sr_{0.27}CuO_4$ | 542[19] | 0.404[14] | | 6.36 $10^{12}$ | 40.5 [&] | 20[19] |
| $YBa_2Cu_4O_{7+y}$ | 140[21] | 0.38[17] | | 8.4 $10^{13}$ | 19.2 [£] | 90[21] |
| BSCCO 2212 | 200[22] | 0.44[18] | | 5.4 $10^{13}$ | 19.2[$] | 82.4[22] |
| Ta | 52[23] | | 5.52 $10^{22}$ [24] | | 560[25] | 4.42[23] |
| Pb | 38[26] | | 1.33 $10^{23}$ [13] | | 200[26] | 7.2[26] |
| Nb | 43[27] | | 5.56 $10^{22}$ [13] | | 35[27] | 9.3[26] |

[#] L is calculated as $\tau v_F$ [28,14]. [&]L is taken the same as for the optimally doped material.

[£]L is calculated as $\tau v_F$ [29,17]. [$]L is calculated as $\tau v_F$[30,18].